\documentclass[a4paper,11pt]{article}
\pdfoutput=1 

\usepackage{jheppub} 

\usepackage[T1]{fontenc} 

\usepackage[dvipsnames]{xcolor}

\def\be{\begin{equation}}
\def\ee{\end{equation}}
\def\bea{\begin{eqnarray}}
\def\eea{\end{eqnarray}}

\def\yzero{\smash{\hbox{$y\kern-4pt\raise1pt\hbox{${}^\circ$}$}}}

\def\beq{\begin{equation}}
\def\eeq{\end{equation}}
\def\beqa{\begin{eqnarray}}
\def\eeqa{\end{eqnarray}}

\def\-{\hphantom{-}}

\def\s2{\frac{1}{\sqrt2}}

\def\beq{\begin{equation}}
\def\eeq{\end{equation}}
\def\beqa{\begin{eqnarray}}
\def\eeqa{\end{eqnarray}}

\def\IF{\relax{\rm I\kern-.18em F}}
\def\II{\relax{\rm I\kern-.18em I}}
\def\IP{\relax{\rm I\kern-.18em P}}
\def\IC{\relax\hbox{\kern.25em$\inbar\kern-.3em{\rm C}$}}
\def\IR{\relax{\rm I\kern-.18em R}}

\def\Dsl{\,\raise.15ex\hbox{/}\mkern-13.5mu D} 
\def\IZ{Z\kern-.4em  Z}

\title{\boldmath The Massive Supermembrane on a Knot}

\author{M.P. Garcia del Moral}
\author{P. Leon}
\author{A. Restuccia\footnote{These authors have contributed equally to this work.}}

\affiliation{Departamento de F\'isica, Universidad de Antofagasta, \\ Aptdo 02800, Chile.}

\emailAdd{maria.garciadelmoral@uantof.cl}
\emailAdd{pablo.leon@ua.cl}
\emailAdd{alvaro.restuccia@uantof.cl}

\abstract{We obtain the Hamiltonian formulation of the 11D Supermembrane theory non-trivially compactified on a twice punctured torus times a 9D Minkowski space-time. It corresponds to a M2-brane formulated in 11D space with ten non-compact dimensions. The critical points like the poles and the zeros of the fields describing the embedding of the Supermembrane in the target space are treated rigorously. The non-trivial compactification generates non-trivial mass terms appearing in the bosonic potential, which dominate the full supersymmetric potential and should render the spectrum of the (regularized) Supermembrane discrete with finite multiplicity. The behaviour of the fields around the punctures generates a cosmological term in the Hamiltonian of the theory.  

The massive supermembrane can also be seen as a nontrivial uplift of a supermembrane torus bundle with parabolic monodromy in $M_9\times T^2$. The moduli of the theory is the one associated with the punctured torus, hence it keeps all the nontriviality of the torus moduli even after the decompactification process to ten noncompact dimensions. The formulation of the theory on a punctured torus bundle is characterized by the $(1,1)-knots$ associated with the monodromies.}

\begin{document}

\maketitle
\flushbottom

\section{Introduction} 

In eleven dimensions there exists a single supergravity theory associated with the low energy limit of  M-theory. When a Kaluza Klein reduction of the eleven dimensional supergravity is done, it leads to the type IIA maximal supergravity in ten dimensions. It is also possible to perform a Scherk Schwarz reduction of the eleven dimensional supergravity. This leads to a ten dimensional  gauged supergravity   \cite{Howe}. In ten dimensions there is also a massive type IIA supergravity found by Romans \cite{romans}.

In \cite{Bergshoeff6} it was shown that it is possible to obtain a non-covariant eleven dimensional supergravity with a cosmological term as an uplift to eleven dimension of the type IIA massive Romans supergravity. The resulting action requires the existence of an isometry in the eleven dimensional space-time, in order to include a non-trivial cosmological massive term. Although the origin of this isometry remains unknown, the proposal for the massive 11D supergravity is a  step forward in understanding the Roman’s supergravity origin.

 It was conjectured by Hull that gauged supergravities are effective limits of M-theory torus bundles with monodromies \cite{Hull4}. At supergravity level, this result was found first  by \cite{Meessen} and later studied in detail in \cite{Bergshoeff5,Melgarejo}. In particular, according to Hull’s conjecture Romans massive supergravity would also admit an M-theory origin. The dual of M-theory compactified on a 2-torus with parabolic monodromy and non-trivially uplifted to ten dimensions could provide an explicit M-theory origin of Romans supergravity  \cite{Hull8}.  This idea was sustained by the fact that  Romans' supergravity, when it is dimensionally reduced, using the Kaluza Klein approach, coincides with the type IIA  gauged supergravity with parabolic monodromy.  Following Hull’s procedure, the authors in \cite{Lowe} proposed a matrix theory formulation for the massive type IIA supergravity.

On the other hand, in \cite{Aharony2} the authors conjectured that the Type IIA string can not be strongly coupled in a weakly curved region of space-time. Their approach was done in the context of supergravity, where the uplift to M-theory is related to the effective field limit of multiple M2-branes with conformal symmetry described by a $U(N)\times U(N)$ Super Chern-Simons matter theory \cite{Bagger,Bagger2,Aharony3, Aharony4}.

One way to study the origin of Roman’s supergravity in M-theory is through the supermembrane theory. We will use interchangeably throughout the work the name supermembrane or M2-brane. The supermembrane action was proposed in \cite{Bergshoeff,Bergshoeff2,Bergshoeff3} and, because of its coupling with the 11D supergravity, it was considered a fundamental object of M-theory. Indeed, through a double dimensional reduction, strings can arise from the membrane. However, in \cite{deWit3} the authors proved that the theory, in a flat space, is unstable at a quantum level. This problem cannot be solved by taking compact dimensions (see \cite{deWit4}). This led the authors in \cite{Banks} to consider that the M2-brane could  be interpreted as a second quantized theory and hence, not be able to describe fundamental degrees of freedom of M-theory. Nevertheless, in \cite{Restuccia}, the authors found a formulation of the M2-brane in $M_9 \times T^2$ with significantly different properties. This formulation, characterized by a restriction over the windings of the theory, known as central charge condition, has a regularized Hamiltonian with  discrete supersymmetric spectrum (see \cite{mpgm12,mpgm11}). This allows us to reinterpret the supermembrane as a fundamental object of a M-theory sector.  This condition has been extended to $4D$ backgrounds, see for example: a $N=1$ toroidally compactified supermembrane with central charges on $M_4\times T^7$  done in \cite{mpgm15}, or on a  $T^7/Z^3_2$ with $G_2$ structure orbifold in \cite{mpgm16}. Another well-known example of  supermembrane with discrete spectrum \cite{mpgm12}, is the supermembrane on a pp-wave background  \cite{Dasgupta,Sugiyama}, whose regularization corresponds to the BMN matrix model \cite{Maldacena3}.

In contrast, the authors of \cite{mpgm2} found that the supermembrane with central charges formulated on $M_9\times T^2$ is the M-theory origin of the type IIB gauged supergravities in 9D. Its Hamiltonian is U-duality invariant, and hence it is also the M-theory origin of type IIA gauged supergravities, in particular those associated to the type IIA parabolic sector \cite{mpgm7}. Globally, the theory is described in terms of a symplectic torus bundle with monodromy in $SL(2,\mathbb{Z})$. Hence, this sector of the M2-brane theory is a natural candidate to realize Hull’s conjecture and model out the M-theory origin of $10D$ Romans' massive supergravity.

Recently, it has been identified  that there is  a one to one correspondence between the M2-brane with central charge condition and the M2-brane with  $C_{\pm}$ fluxes, where the $C_{\pm}$ are the components of the 11D Supergravity three-form in the Light Cone coordinates. Indeed, the supermembrane with central charges in   $M_9 \times T^2$ is equivalent to an M2-brane in $M_9 \times T^2$ with $C_{-}$ fluxes \cite{mpgm6}. Furthermore, the authors showed in \cite{mpgm10} that the M2-brane with $C_{\pm}$ fluxes on a torus bundle theory can be interpreted as an M2-brane formulated on a twisted torus bundle with monodromy in $SL(2,\mathbb{Z})$. The M2-brane theory acquires new  $U(1)$ gauge and global symmetries. There exists a 1-form connection associated with the gauge symmetry $U(1)$, which is dynamical and topologically non-trivial. The supermembrane with $C_{-}$ fluxes formulated on a parabolic torus bundle on $M_9 \times T^2$ can be directly obtained from \cite{mpgm10}, which is based on previous works \cite{mpgm2,mpgm7} and \cite{mpgm6}. 

When a decompactification limit from the M2-brane with $C_-$ fluxes formulated on $M_9\times T^2$ into ten non-compact dimensions is performed, some properties in the uplifted Hamiltonian must be preserved in order to correspond to the massive supergravity sector rather than the massless one. For example, the decompactification limit should preserve finite couplings as well as the moduli of the toroidal structure. These facts cannot be obtained through an ordinary decompactification limit, as it is the case when a 2-torus decompactifies as a cylinder. Indeed,  a $11D$ supermembrane on a cylinder times $9D$ Minkowski target space corresponds to a M2-brane on a circle, i.e. the M2-brane formulated on a trivial circle bundle. Its double dimensional reduction corresponds to the standard $N=2$ 10D type IIA superstring whose effective limit is the $N=2$ 10D type IIA maximal supergravity.

If the Romans massive supergravity has a M2-brane origin, it must arise from a supermembrane defined on a non-trivial bundle in $11D$. There are very few possibilities for implementing this idea with a single compact dimension. Following a former idea of \cite{mpgm13},  we consider the uplift to ten non-compact dimensions in a nontrivial way, $M_9 \times LCD$. The $LCD$ is a Light Cone Diagram, a two dimensional flat strip with identifications and with prescribed segments whose curvature becomes infinite at some points.

Any punctured Riemann surface is equivalent to a LCD diagram. In string theory, the LCD is described by the Euclidean time and a space-like coordinate, the two coordinates of a string, while in our approach it is described by two space-like coordinates. The fields are then functions of three coordinates, the two coordinates of the LCD and the time. So, in the present formulation, the discussion is not in terms of incoming and outgoing strings. It is a more general construction than the interaction string diagrams, as expected since the fundamental objects are now supermembranes. The structure we obtain for the M2-brane on the $M_9\times LCD$ fulfills all of the prescribed requirements: it corresponds to the decompactification limit of a torus bundle times a 9D Minkowski target space into a twice punctured torus bundle times 9D Minkowski space-time, which is effectively a non-trivial compactification with only one compact dimension. Furthermore, it is re-expressed in terms of the LCD, making explicit the tenth noncompact dimension. The connection between the Riemann Surface with punctures and the Mandelstam map describing the LCD plays an essential role in our construction. At supergravity level, the punctures in Riemann surfaces lead to delta function singularities in the equations of motion. In particular, the space-time curvature is singular at the punctures. These singularities can be associated to the existence of Dp-brane sources \cite{Hull10}, where $p$ depends on the dimension of the delta function. In the case of the massive Roman supergravity in ten dimensions and its  uplift to eleven dimensions, it has been related to the  coupling with D8-branes and M9-branes, respectively \cite{Bergshoeff7,Sato}.  There are other examples in the literature of compactifications on Riemann surfaces with punctures  in the context of massive type IIA supergravity, see  \cite{Bah,Bobev,Lozano,Dibitetto} where the punctures are interpreted as the presence of sources generated by Dp-branes or M2-branes stacks.

Indeed, as we will show, our construction corresponds to a massive supermembrane in eleven dimensions, ten of which are non-compact. On top of the massive terms, there is one associated with the punctures that can be understood as a non-vanishing cosmological constant term at the level of supergravity. We expect that the double dimensional reduction of the massive M2-brane naturally leads to the worldsheet action of a type IIA superstring theory with a cosmological term in ten dimensions. This property is expected for the type IIA massive superstring associated to Romans supergravity.  We hope to report about this topic \cite{gmlr2} elsewhere.

On the other hand, from the work \cite{Cattabriga}, it is known that there is a relationship between the $ (1,1)-knots$ and the mapping class group of the twice punctured torus. We will show that the monodromies over the twice punctured torus can be associated with two subgroups of the punctured torus mapping class group, one related to the non-trivial $(1,1)-knots$ and other to trivial knots, being the latter associated to the monodromies of the compact closed torus ( without punctures). We will show that the main properties of the massive M2-brane are identified by the $(1,1)-knots$.

The paper is structured as follows: in section 2, we review the toroidal compactification of the M2-brane on $M_9\times T^2$  with $C_-$ fluxes, or equivalently with central charge, and with a monodromy in $SL(2,\mathbb{Z})$. In section 3, we review some results of the punctured Riemann surfaces that we will extensively use in the next section. In section 4 we obtain the formulation of the supermembrane on a knot, or equivalently on $M_9\times LCD$. We show that the Hamiltonian corresponds to a massive supermembrane with a cosmological mass term. We discuss the amount of supersymmetry preserved  by the theory in this set-up, as well as its global formulation and its relationship to knot theory. In section 5, we present our discussions and conclusions.

\section{The M2-brane with $C_{-}$ fluxes and monodromy in $SL(2,\mathbb{Z})$}\label{sec:MF}
The M2-brane compactified on a torus bundle with $SL(2,\mathbb{Z})$ monodromy is in correspondence, at effective level, with the eight inequivalent classes of  nine dimensions type II gauged supergravities with a monodromy contained in $SL(2,R)$ \cite{mpgm2}. The M2-brane with $C_{\pm}$ fluxes compactified on $M_9\times T^2$ has the same properties. In \cite{mpgm13} it was shown that only M2-brane torus bundles with parabolic monodromy (linearly or non-linearly realized) can be non-trivially uplift to M2-brane bundles on ten non-compact dimensions. The non-triviality of the M2-brane bundle is completely necessary to guarantee that the low energy limit of the decompactified M2-brane corresponds to any of the two massive type IIA supergravities in 10D (Romans and Howe-Lambert-West supergravity) and not to the massless case. 

This is consistent with Hull's conjecture, which stipulated that type IIA massive supergravity can be obtained as the low energy limit of the decompactification of M-theory in a torus bundle with monodromies to ten non-compact dimensions. With this motivation,  it will be useful for the next sections to comment on some of the main results of the torus bundle formulation of the M2-brane with non-trivial monodromies.  We will use some of them to formulate the massive M2-brane on the Light Cone diagram. 

The maps of the  M2-brane (in the light cone gauge L.C.G.) from the base manifold to the target's space torus satisfy the following winding conditions

\begin{eqnarray}
\oint_{C_r}(dX^1+idX^2) &=& 2\pi R(l_s+m_s \tau)\delta^s_r, 
\end{eqnarray}
where $r,s=1,2$. The 2-torus is parametrized in terms of the  moduli $(R, \tau)$ and $(l_s,m_s)$ are the winding numbers of the maps onto the torus in the target space. The winding numbers are organized in the winding matrix $\mathbb{W}=\begin{pmatrix}l_1& l_2\\ m_1& m_2\end{pmatrix}$. $X^1,X^2$ denote the embedding maps of the worldvolume of the supermembrane into the 2-torus $T^2$. The complex closed one-form is denoted by $dX= dX^1+idX^2$. The supermembrane we are interested in, corresponds to an M2-brane with nontrivial central charges i.e. the maps from $\Sigma$ to $T^2$ are restricted by the irreducible wrapping condition

\begin{equation}
\int_{\Sigma} dX^r\wedge dX^s=n \epsilon^{rs} Area(\Sigma), \quad n\ne 0\quad \textrm{and}\quad n\in\mathbb{Z} \label{CH}
\end{equation}
where $\Sigma$ represents the worldvolume of the M2-brane. This is a topological condition that implies the existence of a non-trivial $U(1)$ principal torus bundle with first Chern class $c_1=n$ over the worldvolume of the membrane. On the other hand, this condition implies $det(\mathbb{W})=n$, which is a condition over the winding numbers. The 1-form associated with the embedding maps over the torus can be decomposed in the following way 

\begin{eqnarray}
   dX = dX^1+idX^2 = 2\pi R(l_r+m_r\tau)d\hat{X}^r+dA,  \quad
\end{eqnarray}
where $d\hat{X}^r$ denote the set of normalized harmonic forms over $\Sigma$, that is
\begin{equation}
  \oint_{C_s}d\hat{X}^r = \delta^r_s,
\end{equation}
and $dA=dA^1+idA^2$ is an exact 1-form that transform as a symplectic connection under area preserving diffeomorphisms (APD) connected with the identity.  It is possible to define the determinant worldvolume metric $\sqrt{W}$ as the pull-back of the symplectic 2-form of the target torus 

\begin{equation}
    \frac{1}{2}\epsilon_{ij} d\hat{X}^i\wedge d\hat{X}^j = \frac{1}{2}\epsilon_{ij}\epsilon^{ab}\partial_a \hat{X}^i\partial_b \hat{X}^jd\sigma^1\wedge d\sigma^2 = \sqrt{W}d\sigma^1\wedge d\sigma^2
\end{equation}
being ($\sigma^1,\sigma^2$) local coordinates on $\Sigma$. We label them  by $a,b=1,2$. The central charge condition (\ref{CH}) ensures that $\sqrt{W}$ is always different from zero. Then the Lie bracket is defined as

\begin{equation}
    \{A,B\}=\frac{e^{ab}}{\sqrt{W}}\partial_aA \partial_b B,
\end{equation}
 is also well-defined.

 The Hamiltonian whose compactification on a torus bundle has parabolic monodromy (in the type IIB sector) expressed in complex coordinates, corresponds to,
\begin{equation}
\begin{aligned}
H=&\int_{\sigma}d^2\sigma\sqrt{W(\sigma)}\left[\frac{1}{2}(\frac{P_{m}}{\sqrt{W}})^2+\frac{1}{2}(\frac{P\overline{P}}{W})+\frac{T_M^2}{4}\{X^ {m},X^{n}\}^2+\frac{T_M^2}{2}\mathcal{D}X^m\overline{\mathcal{D}}X^m+\frac{T_M^2}{8}\mathcal{F}\overline{\mathcal{F}}\right]\\
&-\int_{\Sigma} T_M^{2/3}\sqrt{W(\sigma)}\left[\overline{\Psi}\Gamma_{-}\Gamma_{m}\{X^{m},\Psi\}+\frac{1}{2}\overline{\Psi}\Gamma_{-}\overline{\Gamma}\{X,\Psi\}+\frac{1}{2}\overline{\Psi}\Gamma_{-}\Gamma\{\overline{X},\Psi\}\right]\\
& +\int_{\Sigma} \sqrt{W} \lambda\left[ \frac{1}{2}\overline{\mathcal{D}}(\frac{P}{\sqrt{W}})+
\frac{1}{2}\mathcal{D}(\frac{\overline{P}}{\sqrt{W}})+\{X^m,\frac{P_m}{\sqrt{W}}\}-\{\Psi\Gamma_{-},\Psi\}\right],
\end{aligned}
\end{equation}
where $T_{M2}$ is the M2-brane tension. The first class constraint associated with area preserving diffeomorphisms has been incorporated into the Hamiltonian through a Lagrange multiplier $\lambda$.
The symplectic covariant derivative is defined as,

\begin{equation}
\mathcal{D} \bullet=D\bullet+\{A,\bullet\},\quad \mathcal{F}=D\overline{A}-\overline{D}A+\{A, \overline{A}\}
\end{equation}
with $D=D_1+iD_2$ and $D_r$ defined as follows,

\begin{equation}
D_r\bullet=\frac{\epsilon^{ab}}{\sqrt{W}}2\pi R (l_r+m_r\tau)\theta_r^s\partial_a\widehat{X}^s \partial_b\bullet,
\end{equation}
and with $\theta\in SL(2,\mathbb{Z})$ a matrix defined in terms of the monodromy matrix $\rho$.

The effect of imposing the central charge condition on the toroidally compactified supermembrane changes qualitatively the behaviour of the mass spectrum. It becomes purely discrete \cite{mpgm,mpgm12,mpgm11} in contrast with the general case, not restricted by the topological condition (\ref{CH}), where the spectrum is continuous from $[0,\infty)$ . Equivalently, the same result holds if we consider a supermembrane  compactified on the same background but now subject to the effect of $C_-$ fluxes described by 
\bea
\int_{T^2}C_{-}=n \label{qcf}
\eea
acting on the target space 2-torus, and $n$ representing the units of flux. As shown in \cite{mpgm6}, the condition (\ref{qcf}) is a quantization condition over the components of the 3-form background $C$. Moreover, the pull-back of Eq. (\ref{qcf}) to the base manifold implies the central charge condition over the worldvolume surface.

It is important to mention that under the equivalent conditions (\ref{CH}) or (\ref{qcf}) the torus embedding becomes irreducible and it cannot degenerate. The theory contains pure supermembrane excitations - the fields depend on the three worldvolume coordinates-  as well as string-like configurations which carry non-trivial energy as in string theory, in distinction to the zero energy string spikes which are present in the $11D$ supermembrane as discussed in [19]. A string theory is obtained after a double-dimensional reduction.  To do so, one has to freeze part of the supermembrane degrees of freedom, in the sense that the configurations depend only on one spacelike coordinate. In this work we will not consider its double dimensional reduction.

Globally, the M2-brane in any of the two cases  discussed above, corresponds to a supermembrane realized on symplectic torus bundle with monodromy whose inequivalent classes are  specified by the second cohomological class $H^2(\Sigma,\mathbb{Z}_{\rho_p})$ \cite{mpgm2,mpgm3}.

On the worldvolume base manifold $\Sigma$, there are two non-trivial bundles: the non-trivial $U(1)$ principal bundle, characterized by the first Chern class $c_1=n$, and the symplectic torus bundle with monodromy. Both bundles define together a twisted torus bundle over the base $\Sigma$ on which the M2-brane maps are defined as sections \cite{mpgm10}. This condition restricts the type of M2-brane torus bundles that can be defined in $M_9 \times T^2$. They lie in two different classes: eight characterized by the  non-vanishing topological numbers $(n,\rho)$ associated to the M2-brane sector with central charge with discrete spectrum, and another one with no topological charges  $(0,0)$ associated to a trivially wrapped M2-brane with continuous spectrum. At low energies, they are associated with the eight inequivalent classes of type IIA  gauged supergravities and with the type IIA maximal supergravity in 9D, respectively. 
 
The Hamiltonian of the Supermembrane considered here is locally and globally U-duality invariant  as shown in the paper \cite{mpgm7}. It also possesses a residual global symmetry $\rho_p\times \rho_{p}^*$ contained in $SL_{\Sigma^2}(2,\mathbb{Z})\times SL_{T^2}(2,\mathbb{Z})$ associated with the monodromy class of matrices of the fiber and on the base manifold.

 The parabolic monodromy, in contrast with other monodromies contained in $SL(2,\mathbb{Z})$, allows a decompactification procedure to ten non-compact dimensions while keeping a non-trivial topology. In 10D there are only two massive deformations of type IIA supergravity (HLW and Romans supergravities). Under reduction to nine dimensions, they generate a gauged supergravity with parabolic and a gauged trombone monodromy, respectively. In the M2-brane torus bundle classifications among others, there are also two inequivalent class of bundles with parabolic monodromies, one linearly and another nonlinearly realized. The second one is associated to a "trombone" monodromy in correlation with its low energy description. Under decompactification, the formerly M2-brane with a parabolic monodromy linearly realized will be formulated  on a twice punctured torus target space with monodromy. The formulation on a punctured surface will lead to an eleven dimensional  theory with only one compact dimension. In section 4. we will show a realization of this idea.
\section{Parametrization of the twice punctured torus}

A previous step to obtain the formulation of the M2-brane on a punctured torus is to define the parametrization of the space and embedding maps on this Riemann surface. We will first review some basic results about the Riemann surfaces with punctures. It is known from \cite{Mandelstam,Mandelstam2,Giddings2,Ito}  that the $N$ punctured Riemann surfaces of genus $g$ are conformally equivalent to a $g$ loop Light Cone Diagrams (LCD) -string interaction-. Specifically, given an abelian differential dF on an arbitrary Riemann surface with punctures, $\Sigma_{g,N}$, it is always possible to define a flat metric with isolated singularities representing the punctures -incoming/outgoing states (strings)-. This equivalence has been exploited in a different context to  compute scattering amplitudes in string theory, see for example \cite{SONODA,Restuccia5,Ishibashi4}. As already explained in the introduction, our construction is more general  than the interaction string diagrams, as expected since the fundamental objects are now supermembranes.

Defining a complex coordinate system $z$ over $\Sigma_{g,N}$ the conformal map (Mandelstam map) can be written as

\begin{equation}
 F(z)=\int^z_{z_0} dF = \sum_r \alpha_r\left[ln E(z,Z_r)-2\pi i \int_{z_0}^{z} w \frac{1}{Im\Omega} Im\int_{z_0}^{Z_r} w\right],
\end{equation}     
 where $z_0$ is an arbitrary point on $\Sigma_{g,N}$, $Z_r$ with $ (r=0..N)$ denote the positions of the punctures on $\Sigma_{g,N}$. On the LCD, $\alpha_r$ are  the weights associated to the punctures, which satisfy $\sum \alpha_r=0$ and $w=(w)_j$ with $(j=1,...,g)$ is the basis of holomorphic 1-forms, on the compact without punctures Riemann surface, normalized as

\begin{equation}
\int_{a_j} w_i = \delta_{ij}, \quad \int_{b_j}w_i = \Omega_{ij}.
\end{equation}  
where $a_j$,$b_j$ are are the basis of the Riemann surface's homology cycles, and $\Omega_{ij}$ is the period matrix. Finally, $E(z,Z_r)$ it is a prime form defined for an odd spin structure and in terms of the theta functions with characteristic $[s]$ as
 
\begin{equation}
 E(z,Z_r)= \frac{\Theta[s](\int_{Z_r}^z w,\Omega)}{h_s(z)h_s(Z_r)},
\end{equation}  
with 

\begin{equation}
h_s(z)= \sqrt{\sum_j \frac{\partial \Theta[s]}{\partial \xi_j}(0,\Omega)w_j(z)},
\end{equation}
where the notation $[s]$ is used to represent, in general, a point $[s]=\left[\begin{array}{c} u\\v \end{array}\right]\in \mathbb{C}^g$ with $u,v\in\mathbb{R}^g$. Now, if $u,v\in \left(\mathbb{Z}/2\mathbb{Z}\right)^g$, [s] receive the name of spin structure and it is said to be odd or even if $uv$ is odd or even (see for example \cite{Fay,mumford}).

The 1-form $dF$ has poles at the punctures and also have zeros at $P_a$ with $a=0,..,2g-2+N$ , whose position can be computed in terms of the moduli of the Riemann surface (see for example  \cite{Ito}). The set of parameters that characterize the Riemann Surface $\Sigma_{g,N}$ are the Teichm\"uller parameter $\tau$ and the positions of the punctures $Z_r$. On the other hand, the set of parameters that describe the LCD are the external weights  $\alpha_r$, the internal weights $\beta_j$,  the internal lengths $T_u$, with $(u=1,..,2g-3+N)$, and the twist angles $\theta_v$ with $(v=1,..,3g+N-3)$. However, not all these parameters are independent. In fact,  the following relationship between the moduli of both surfaces can be found (see \cite{Giddings2} for more details) 

\begin{equation}
\int_{b_j} dF - \sum_{i}^g \Omega_{ij}\int_{a_j}dF=2\pi i \int_{Z_2}^{Z_1}w_j.   \label{ide} 
\end{equation}

The case of interest for us corresponds to a Riemann surface with $g=1$. The associated basis of the holomorphic differential has only one element,
\bea
dz=w,
\eea
that satisfies
\bea
\oint_aw=1,\quad \oint_b w=\tau,
\eea
being $\tau$ the Teichmuller parameter of the 2-torus. Thus  $F(z)$, for $N=2$ punctures with residues $\alpha_r=(-1)^{r+1}\alpha$ can be written as  

\bea
F(z)=\alpha \ln\left[\frac{\Theta_1(z-Z_1\vert \tau)}{\Theta_1(z-Z_2\vert \tau)}\right]-2\pi i \alpha \frac{Im (Z_1-Z_2)}{Im \tau}(z-z_0).
\eea

\begin{figure}
    \centering
    \includegraphics[scale=0.4]{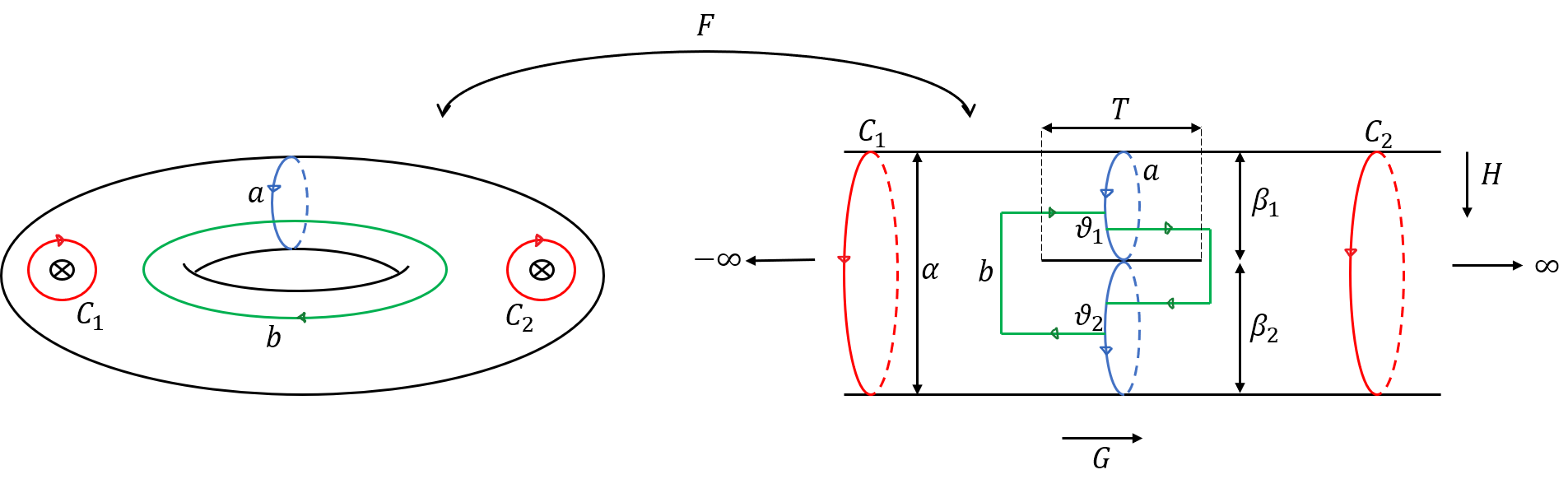}
    \caption{The torus with two punctures and the corresponding LCD. The Mandelstam map send the punctures over the torus to $\pm\infty$ in the LCD.}
    \label{fig:equi}
\end{figure}

Now, we map the torus with two punctures on the one loop light cone diagram.  Apart from the definition of the map $F(z)$,  the following identification holds (\ref{ide})

\begin{equation}
2\pi i (Z_1-Z_2)= (\theta_1+\theta_2)\beta_1-\theta_2-2\pi i \beta_1 \tau
\end{equation}  
where the moduli of the twice punctured torus are characterized by the four parameters $(\tau,Z_1,Z_2)$ and the moduli of the one loop light cone diagram are described by $(\alpha,\beta_1,\theta_1,\theta_2)$, see figure (\ref{fig:equi}). These quantities can be defined in terms of integrals of $dF$ as

\begin{eqnarray}
     \int_{a}dF= 2\pi i \beta_1, \quad \int_{b}dF = \frac{i}{2\pi}(\beta_1\theta_1-\beta_2\theta_2),\quad \int_{C_r}dF=(-1)^r2\pi i \alpha, \quad T=\int_{P_1}^{P_2}dF. \nonumber \\  \label{tA}
\end{eqnarray}
with $C_r$ curves around each of the two punctures and $\beta_2=\alpha-\beta_1$. Finally, it will be useful for the following sections to express the map $F(z)$ as 

\begin{equation}
F= G+iH
\end{equation}
where $G=Re(F)$  corresponds to the Green function, with $G$  single-valued, that is , 

\begin{equation}
\int_
{a}dG=\int_
{b}dG=0
\end{equation}
At the same time, the  associated 1-form  $dG$ is harmonic since it has poles at the punctures such that the sum of the residues over the compact manifold is zero. Around the punctures, the function $G$ satisfies,
\begin{eqnarray}
G \sim  (-1)^{r+1}\alpha\ln\vert z-Z_r\vert,
\end{eqnarray}
 $H= Im(F)$ is multivalued and $dH$ is harmonic. Usually the function $G$ is interpreted as the light cone time in the context of strings interaction and $H$ behaves locally, near each puncture as an angle  

\begin{equation}
dH = (-1)^{r+1}\alpha d\varphi, \quad \mbox{with} \quad \varphi\in (0,2\pi) \ \ (r=1,2). \label{HB}
\end{equation}
On the other hand, close to the zeros of $dF$ denoted as $P_a$, the functions $G$ and $H$ can be written as
\begin{eqnarray}
G(z)-G(P_a) &\sim &  \frac{1}{2}Re(D(P_a)(z-P_a)^2),\label{Gz}  \\
H(z)-H(P_a)  & \sim &  \frac{1}{2}Im(D(P_a)(z-P_a)^2), \label{Hz}
\end{eqnarray}
where
\begin{equation}
   D(P_a) = \sum_{r=1}^{2} (-1)^{r+1} \left[ \frac{\partial^2_z \Theta_1(P_a-z_r,\tau)}{\Theta_1(P_a-z_r,\tau)}-\left(\frac{\partial_z \Theta_1(P_a-z_r,\tau)}{\Theta_1(P_a-z_r,\tau)}\right)^2 \right].
\end{equation}

\begin{figure}
    \centering
    \includegraphics[scale=0.4]{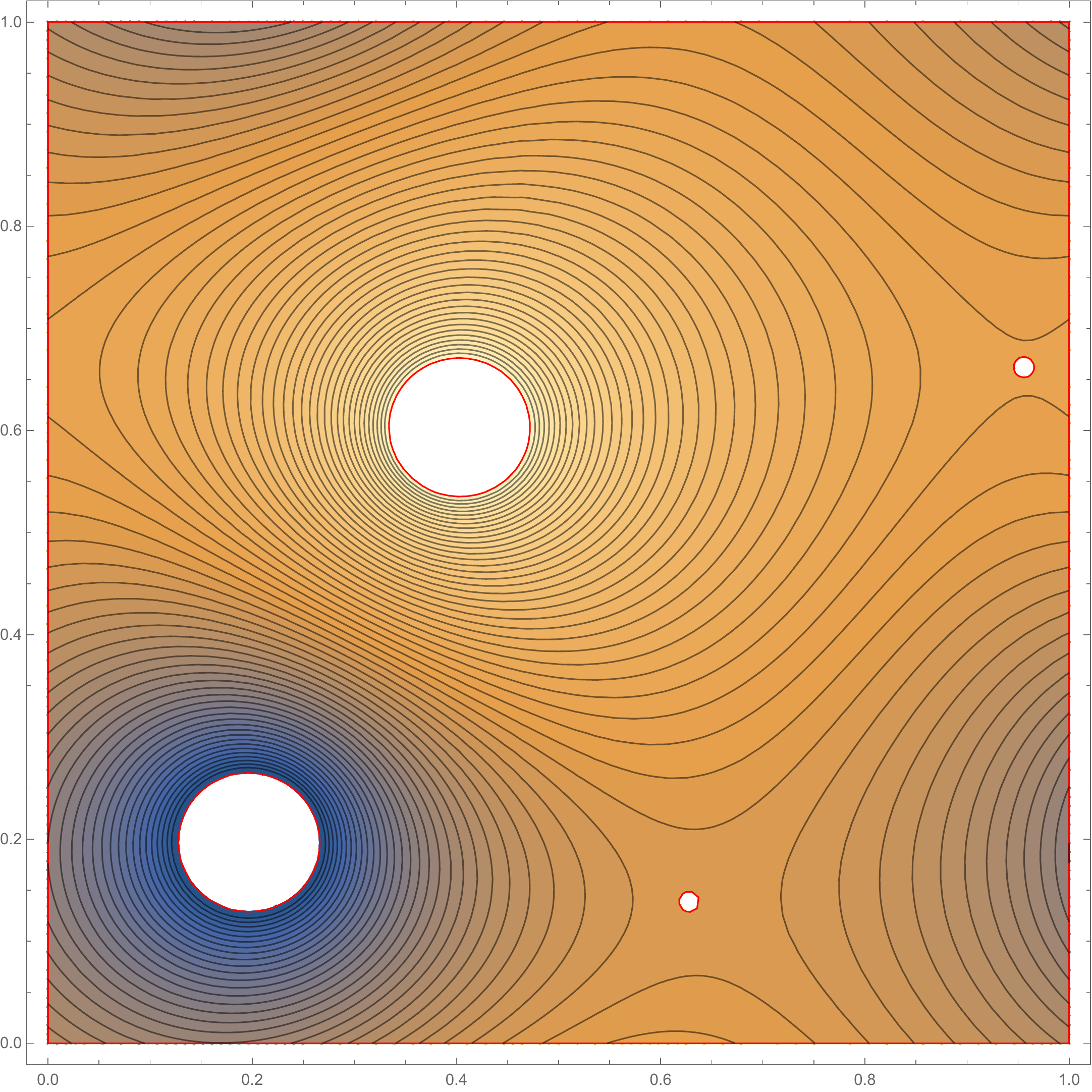}
    \caption{Curves $G=const$ with $\tau=i$, $Z_1=0.2(1+i)$, $Z_2=0.4+0.6i$, $P_1=0.643+0.129i$ and $P_2=0.956+0.662i$. The big and small white circles represent the positions of the punctures and the zeros of $dF$, respectively. }
    \label{fig:G}
\end{figure}

\begin{figure}
    \centering
    \includegraphics[scale=0.4]{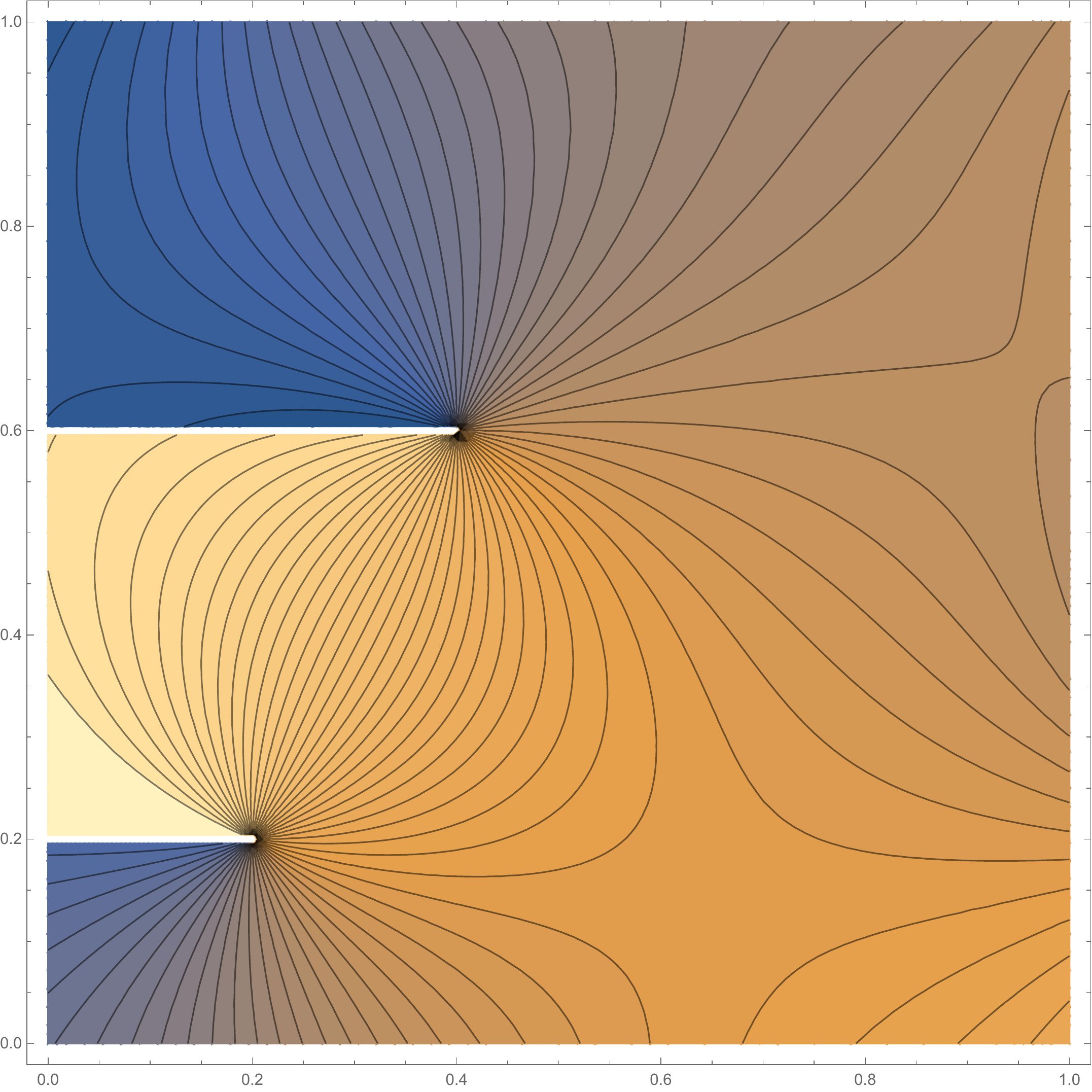}
    \caption{Curves $H=const$ with $\tau=i$, with $\tau=i$, $Z_1=0.2(1+i)$, $Z_2=0.4+0.6i$ $P_1=0.643+0.129i$ and $P_2=0.956+0.662i$.}
    \label{fig:H}
\end{figure}

The behavior of the functions $G$ and $H$ become clearer if the Riemann surface is modeled as a polygon and we plot the curves $G=const$ and $H=const$, see figures (\ref{fig:G}) and (\ref{fig:H}). In these graphics, it turns out that the curves $G=const$ start as closed curves around one of the punctures until they reach one of the zeros of $dF$ where the curve breaks into two closed curves (each one homotopically equivalent to the curve $a$). Then, in the second zero of $dF$, the curves come together again and form a closed curve around the second puncture.

On the other hand, the curves $H=const$ are curves that go from one puncture to the other. The lines of $H=const$ shown in the diagrams that seem not to connect the two punctures are curves defined outside the fundamental domain of the Riemann surface that equally end in punctures shifted by the periodicity of the lattice, equivalent to those defined inside the fundamental domain.

\section{The M2-brane action on a twice punctured torus}
In this section, we will study the decompactification limit of the M2-brane with fluxes presented in section (\ref{sec:MF}) to a M2-brane formulated on an eleven dimensional target space with ten non-compact dimensions. To that end, we propose to formulate the M2-brane on a background which is given locally by $M_9\times LCD$, where $LCD$ represents a more convenient surface for modeling out the twice punctured torus. Every punctured Riemann surface is conformally equivalent to a string L.C interaction diagram.  

The metric over LCD that we use corresponds to 
\begin{eqnarray}
     ds^2 &=& \frac{T^2}{\cosh^4{\hat{G}}}d\hat{G}^2+dH^2 \label{mono0},
\end{eqnarray}
where $\hat{G}\equiv G/\alpha \in(-\infty,\infty)$ and $T\equiv G(P_2)-G(P_2)$ represents the internal length of the LCD, see figure \ref{fig:equi}. Now, it will be useful to define the variable $K\equiv T^2 \tanh{\hat{G}}$ which maps the real line into the interval $[-T,T]$. Then the metric can be rewritten as the following almost flat metric

\begin{eqnarray}
     ds^2 &=& dK^2+dH^2 \label{mono2}.
\end{eqnarray}
It is important to mention that this  metric has a curvature different from zero at the zeros and poles of $dF$. A comment is in order, $X^m, G,H$ with $m=1,\dots,7$ are the transverse LC target space coordinates. On the LCD, the punctures associated with $G$ are mapped to $\pm\infty$. They are arbitrarily located over the Riemann surface. The area preserving diffeomorphisms on the Riemann surface are restricted to the ones which preserve the punctures and hence the zeros of the holomorphic 1-form dF. 

Now, we consider the decompactification limit to ten non-compact dimensions of the supermembrane with central charges (or equivalently with $C_-$ fluxes)\cite{mpgm6}, formulated on a torus bundle with parabolic monodromy \cite{mpgm3,mpgm6}). It is convenient to define the punctured Riemann surface, where the associated zeros have also been extracted. We define the Riemann surface has having neighbourhoods U1, U2 around the poles of F(z) and U3, U4 around the zeros of F(z), as well and their homeomorphisms, that are compatible with the Riemann surface's holomorphic structure, which maps each neighborhood  into a punctured disk on $\mathbb{R}^2$. We define the determinant of the worldvolume metric $\sqrt{W}$ on $\Sigma_{1,2}$ as the pull-back of the volume 2-form on $LCD$,

\begin{eqnarray}
     \sqrt{W} &=& \epsilon^{ab} \partial_a K\partial_b H.
\end{eqnarray} 
In this way the Lie bracket of the maps $K$ and $H$ is well-defined, 
\begin{equation}
\{K,H\}=1 .
\end{equation}
Notice that this bracket can be extended to the zeros of $dF$. 

In order to write the decompactified Hamiltonian we compare the construction with the one in \cite{mpgm6}. In the reference there is a flat 2-torus $T^2$ on the target, a torus $\Sigma$ as a base manifold with a symplectic structure induced from the canonical one on the target. The one-to-one maps from $\Sigma$ to the target $T^2$ are built using harmonic 1-forms on $\Sigma$ defined as the real and imaginary parts of the holomorphic 1-form defined on $\Sigma$. They are  closed forms whose harmonic parts are the ones defined previously, plus an exact 1-form which carries the physical degrees of freedom.  The above construction is restricted by the topological central charge or flux condition, which is associated with the existence of a rich structure of connections in non-trivial bundles. 

Once decompactified, there  exists new closed non-trivial 1-forms
$$dX^K= dK+ dA^{K},\quad dX^H= mdH+ dA^{H},$$
where $m$ is an integer that near the puncture can be interpreted as a winding number. Assuming that the fields $X^m$, $A^K$, $A^H$ and $\Psi$  are well defined on the compact surface, including the points at which $dF$ has poles or zeros, then the Lagrangian density can be written as

\begin{eqnarray}
   \mathcal{L} &=& T_M\frac{\sqrt{W}}{2}\Bigg[(\dot{X}^m)^2+(\dot{X}^K)^2+(\dot{X}^H)^2+\bar{\Psi}\Gamma^-\dot{\Psi}+\frac{1}{2}\{X^m,X^n\}^2+\{X^K,X^n\}^2 \nonumber \\  &+& \{X^H,X^n\}^2+\{X^K,X^H\}^2 + 2\bar{\Psi}\Gamma^-\Gamma_m \{X^m,\Psi\}+2\bar{\Psi}\Gamma^-\Gamma_K \{X^K,\Psi\} \nonumber \\ &+& 2\bar{\Psi}\Gamma^-\Gamma_H \{X^H,\Psi\} \Bigg].
\end{eqnarray}
 The conjugate momenta are given by

\begin{eqnarray}
   P_m & = &\frac{\partial \mathcal{L} }{\partial \dot{X}^m} = T_M\sqrt{W} \dot{X}^m, \\
   P_K & = & \frac{\partial \mathcal{L} }{\partial \dot{X}^K} = T_M\sqrt{W} \dot{X}^K, \\
   P_H & = & \frac{\partial \mathcal{L} }{\partial \dot{X}^H} = T_M\sqrt{W} \dot{X}^H, \\
   S & = & \frac{\partial \mathcal{L} }{\partial \dot{\bar{\Psi}}} = -T_M\sqrt{W}\Gamma^-\Psi,
\end{eqnarray}
where it is clear that $P_m/\sqrt{W},P_K/\sqrt{W},P_H/\sqrt{W}$ and $S\sqrt{W}$ are also well defined over the compact surface. 

in the definition of the maps from $\Sigma_{1,2}$  to the target space, we introduce the harmonic one-form $d ( G+iH)$ rather than the harmonic one-form $d\hat{X}$.
Locally, $dG$ and $dH$, both contribute to the local constraint at the same level,

\bea
\{P_{K},X^K\}+\{P_H,X^H\}+\{P_m,X^m\}=0
\eea
but globally, they have very different properties. On $\Sigma_{1,2}$, $G$ is a single-valued function whereas $H$ behaves like an angle coordinate around the punctures. They define a well-behaved coordinate system on the surface, away from the punctures and the zeros of $F(z)$. 

 Since $X^{K}$ and $K$ are scalars on $\Sigma_{1,2}$  then necessarily $A^{K}$ is also a scalar. Hence, $A^{K}$ does not transform as the component of a symplectic connection under symplectomorphism transformations, as it happens in the original theory before the uplifting, that is, in the M2-brane with central charges on $M_9\times T^2$ analyzed in \cite{Ovalle1, mpgm10}. Consequently, $(A^K,A^H)$ cannot be interpreted as the components of a connection. The reason of this difference lies in the target space geometry considered, since there is only a single compact dimension. 

The moduli of $\Sigma_{1,2}$ as well as the parabolic monodromy are introduced in the uplifted  Hamiltonian through the Mandelstam map. The functions $G$ and $H$ depend only on the Riemann surface and the punctures on it, they are not physical degrees of freedom. The maps defining the Supermembrane take value on  a eleven dimensional space with ten non-compact dimensions. The image of the map  $G$ from  the Riemann surface on the real line $\mathbb{R}^1$ is the 10th non-compact coordinate. $G$ on the Riemann surface is a height function with values ranging from $-\infty$ on one puncture to $+\infty$ on the other puncture.  The 11th coordinate takes values on the compact sector of the target space. Indeed, its global description represents a fibration of the compact dimension described by $H$ on $\mathbb{R}$. 

\begin{figure}
    \centering
    \includegraphics[scale=0.5]{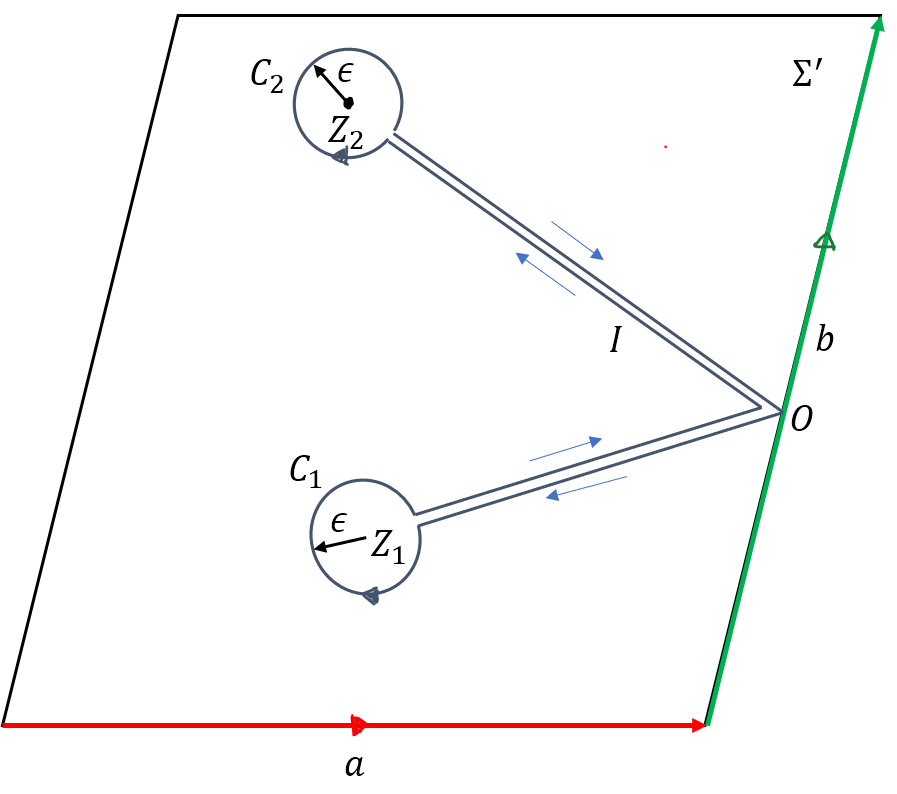}
    \caption{The region $\Sigma'$ obtained by cutting $\mathbf{\Sigma}$ through the curves $C_1$,$C_2$ and $I$. The path obtained by the union of the curves $C_1$,$I$,$C_2$ and $I^{-1}$ is denoted by $c$}
    \label{fig:SigmaP}
\end{figure}

In order to write the Hamiltonian of the M2-brane,  we need to define how to perform the integration over $\Sigma_{1,2}$. This is necessary since the metric over the base manifold is defined in terms of the real and imaginary parts of the Mandelstam map and consequently it may exhibit problems at the singular points, this is, the punctures and zeros of $dF$. Firstly let us notice that the set of zeros associated to $dF$, $S\equiv(P_1,P_2)$, it has null measure. Therefore, by assuming that the Hamiltonian density is well-defined on $S$, we have

\begin{equation}
    H=\int_{\Sigma_{1,2}}\mathcal{H}=\int_{\Sigma_{1,2}/S}\mathcal{H},
\end{equation}
which indicate that, at least from the Hamiltonian point of view, the only problematic points are the punctures. Then, to define the integral taking into account these points (see \cite{farkas}), first consider the region $\Sigma_{1,2}$ as the fundamental domain of the 2-torus $\mathbf{\Sigma}_{1,2}$, defined as the quotient of $\mathbb{R}^2/\Gamma$ were $\Gamma$ is a discrete group with generators 1 and $\tau$.\newline 
To extract the punctures in such a way that the resulting region will be simply connected, we will cut the fundamental region from a point $O\in \partial\mathbf{\Sigma}_{1,2}$ to one of the punctures to return again to $O$ and then go to the second puncture to finally go back to $O$. As a result, the resulting region $\Sigma'$ does not contain any puncture and is simply connected, see figure \ref{fig:SigmaP}.

Now the Hamiltonian of the M2-brane in the L.C.G formulated on the LC diagram corresponds to

\begin{eqnarray}\label{Hamiltonian massive}
     H &=& \frac{1}{2T_M} \lim_{\epsilon\rightarrow 0} \int_{\Sigma'} d^2\sigma \sqrt{W} \bigg[ \left(\frac{P_m}{\sqrt{W}}\right)^2+\left(\frac{P_{K}}{\sqrt{W}}\right)^2+\left(\frac{P_{H}}{\sqrt{W}}\right)^2+\frac{T_M^2}{2}\{X^m,X^{n}\}^2 \nonumber\\[0.3cm] &+& T^2_M \{K+A^{K},X^{n}\}^2
 +T^2_M\{mH+A^{H},X^{n}\}^2+ T^2_M\{K+A^{K},mH+A^{H}\}^2   \nonumber \\[0.3cm] &+&  2T^2_M\bar{\Psi}\Gamma^-\Gamma_m \{X^m,\Psi\}+2T^2_M\bar{\Psi}\Gamma^-\Gamma_K \{X^K,\Psi\}+2T^2_M\bar{\Psi}\Gamma^-\Gamma_H \{X^H,\Psi\} \bigg],
\end{eqnarray}
which is subject to the following local constraint associated to the Area Preserving Diffeomorphims (APD)
  \begin{equation}
     \phi_1=d\left[\left(\frac{P_K}{\sqrt{W}}\right)dX^K+\left(\frac{P_H}{\sqrt{W}}\right)dX^H+\left(\frac{P_m}{\sqrt{W}}\right)dX^m + \bar{\Psi}\Gamma^-d\Psi \right]=0 \label{lc}
 \end{equation}
and three global constraints 
\begin{eqnarray}
     \zeta _r &=&\int_{\mathcal{C}_r}\left[\left(\frac{P_K}{\sqrt{W}}\right)dX^K+\left(\frac{P_H}{\sqrt{W}}\right)dX^H+\left(\frac{P_m}{\sqrt{W}}\right)dX^m + \bar{\Psi}\Gamma^-d\Psi \right]=0, \quad r=1,2\label{g1} \nonumber \\ && \\
   \zeta_3 &=&\int_{C_1}\left[\left(\frac{P_K}{\sqrt{W}}\right)dX^K+\left(\frac{P_H}{\sqrt{W}}\right)dX^H+\left(\frac{P_m}{\sqrt{W}}\right)dX^m + \bar{\Psi}\Gamma^-d\Psi \right]=0, \label{g3}
 \end{eqnarray}
where the first two constraints are the usual ones,  $\mathcal{C}_r$, $r=1,2$ is the basis of homology cycles on $\Sigma$ (we are using $\mathcal{C}_1=a$ and $\mathcal{C}_2=b$)  and the extra constraint appears due to the presence of singularities associated with the punctures. There is no need to introduce a fourth constraint associated with the other puncture since it is not an independent one. It can be shown that it corresponds to a linear combination of the preceding three global constraints (see \cite{farkas}).  
\subsection{A Massive M2-brane formulated on ten non-compact dimensions} 
We want to prove that the $11D$ supermembrane compactified on a target space locally given $M_9\times LCD$ with the metric (\ref{mono2}) is a massive M2-brane formulated on ten non-compact dimensions. In order to do this we will show that the quadratic contribution present in the bosonic potential of the Hamiltonian is non-vanishing. The bosonic potential is given by, 
\begin{eqnarray}
     V_B &=& \{K+A^K,X^m\}^2+\{mH+A^H, X^m\}^2+\{X^m,X^n\}^2 +\{K+A^K, mH+A^H\}^2. \nonumber \\ && \label{bp}
\end{eqnarray}
Thus, the quadratic terms of the dynamics field contained in the potential are
\begin{equation}
   \mathcal{V} = \{K,X^m\}^2+m^2\{H,X^m\}^2+\{K,A^H\}^2+m^2\{H, A^K\}^2+m\{K,A^H\}\{A^K, H\}
\end{equation}
It may seem that they do not generate mass terms for every field. Nevertheless, the mass terms become manifest if we impose the following gauge fixing, which is analogue to the Coulomb gauge,
\bea
\{K,A^K\}+m\{H,A^H\}=0. \label{gc}
\eea 
Indeed, since this expression is identically  zero, we can square it and add it to $\mathcal{V}$ and obtain the following result for the quadratic terms 
\begin{eqnarray}
\mathcal{V} & = & \{K,X^m\}^2+m^2\{H,X^m\}^2+  \{K,A^K\}^2+m^2\{H,A^K\}^2   \nonumber \\ &+&\{K,A^H\}^2 + m^2\{H,A^H\}^2 
\end{eqnarray}
We can observe that, besides the quartic potential, there have also been generated quadratic terms for all the fields. The final step  is to prove that the quadratic terms associated with the non-trivial forms $dK,dH$ of the target space, are always different from zero in $\Sigma/S$ as long as the dynamic fields  $X^m, A^K, A^H$ are not-trivial and $T\not=0$,$m\not=0$, i.e.
\begin{eqnarray}
\{K, X^m\}^2+m^2\{H,X^{m}\}^2 &\ne& 0,  \label{mass1}\\
\{K,A^K\}^2+m^2\{H,A^K\}^2 &\ne&  0, \label{mass2} \\ 
\{K,A^H\}^2+m^2\{H,A^H\}^2 &\ne&  0. \label{mass3}
\end{eqnarray}

\subparagraph{Proof} In the membrane theory, trivial fields imply that they are constant on the punctured Riemann surface. In fact, the field always appears in the Hamiltonian as one forms on the surface, hence they are defined modulo constants on the surface.  We thus assume that the fields are not constant on the Riemann surface. We will prove by contradiction.
 Let us begin considering the Eq. (\ref{mass1}). Suppose that for each point in $\Sigma'/S$
 \begin{equation}
    \{K, X^m\}^2+m^2\{H,X^{m}\}^2 = 0 \Rightarrow \{K, X^m\}=\{H,X^{m}\}=0,  \label{ab}
 \end{equation}
which  can be rewritten as
\begin{eqnarray}
    \partial_1 K \partial_2 X^n -  \partial_2 K \partial_2 X^n & = & 0, \\
    \partial_1 H \partial_2 X^n -  \partial_2 H \partial_2 X^n & = & 0.
\end{eqnarray} 

By assumption $X^m$ is not constant on the Riemann surface. On the surface, there exists then some open set where at least one  derivative of $X^m$ is not zero. On that set we have
\begin{eqnarray}
\frac{\partial_1G}{\partial_2 G}=\frac{\partial_1X^m}{\partial_2 X^m}, \quad \frac{\partial_1H}{\partial_2 H}=\frac{\partial_1X^m}{\partial_2 X^m},
\end{eqnarray}
which implies
\begin{equation}
\frac{\partial_1G}{\partial_2 G}= \frac{\partial_1H}{\partial_2 H},
\end{equation}
but $F=G+iH$ is a holomorphic function in $\Sigma'$ and consequently
\begin{equation}
\partial_2 G=-\partial_1 H, \quad  \partial_1 G=\partial_2 H,
\end{equation}
 so $(\partial_2 G)^2+(\partial_1 G)^2=0$ which is a contradiction since $G$ and $H$ are not constant. It easy to see that this statement also holds for (\ref{mass2}), (\ref{mass3}) and consequently all of the mass terms must be different from zero.

Another interesting feature that is only present in this formulation comes from the last term of the bosonic potential (\ref{bp}), specifically

\begin{eqnarray}
  && \lim_{\epsilon\rightarrow 0}\int_{\Sigma'}d^2 \sigma\sqrt{W}\{K,H\}\left(  m\{K,A^H\}+m^2\{H,A^K\}+\frac{m^2}{2}\{K,H\}\right)  \nonumber \\  && \hspace{4cm} = \lim_{\epsilon\rightarrow 0}\int_{\Sigma'} \left( mdK\wedge dA^H + m^2dA^K\wedge dH+\frac{m^2 }{2}dK\wedge dH \right). \nonumber \\ &&
\end{eqnarray}
The first integral can be written as 
\begin{eqnarray}
   \lim_{\epsilon\rightarrow 0}\int_{\Sigma'} dK\wedge dA^H &=& - \lim_{\epsilon\rightarrow 0}\int_{\partial\Sigma'}A^HdK= - \lim_{\epsilon\rightarrow 0}\left[\int_{C_1}+\int_{C_2}+\int_{I}+\int_{I^{-1}}\right]A^HdK=0, \nonumber \\ && 
\end{eqnarray}
while the second integral 
\begin{eqnarray}
   \lim_{\epsilon\rightarrow 0}\int_{\Sigma'}dA^K\wedge dH & = &   \lim_{\epsilon\rightarrow  0}\left[\int_{C_1}+\int_{C_2}+\int_{I}+\int_{I^{-1}}\right]A^KdH \nonumber  \\ &=& \frac{1}{2i}\lim_{\epsilon\rightarrow  0} \sum_r \int_{C_r}A^K(dF-d\bar{F}) = -\alpha \lim_{\epsilon\rightarrow  0} \sum_r \int_{0}^{2\pi}(-1)^{r}A^K(Z_r+\epsilon e^{-i\theta})d\theta \nonumber \\ &=& 2\pi \alpha [A^K(Z_2)-A^K(Z_1)]. 
\end{eqnarray}
Finally the last integral leads to 

\begin{eqnarray*}
\frac{1}{2}\lim_{\epsilon\rightarrow 0}\int_{\Sigma'}dK\wedge dH & = &   \frac{1}{2}\lim_{\epsilon\rightarrow  0}\left[\int_{C_1}+\int_{C_2}+\int_{I}+\int_{I^{-1}}\right]KdH =  \lim_{\epsilon\rightarrow  0} K_r \sum_r \int_{C_r} dH,
\end{eqnarray*}
where we use the fact that $C_r$ are curves $K=const$, specifically

\begin{eqnarray}
  K(Z_r) = (-1)^{r+1}T\left(\frac{\epsilon^{2\alpha}-1}{\epsilon^{2\alpha}+1}\right).
\end{eqnarray}
Then, using (\ref{HB}), we get
\begin{eqnarray} 
   \frac{1}{2}\lim_{\epsilon\rightarrow 0}\int_{\Sigma'}dK\wedge dH = 2\pi \alpha T. \label{fc}
\end{eqnarray}

Now, using the definition the canonical momenta zero modes,
\begin{equation}
    P_{0H}=\int_{\Sigma^{'}}d\sigma^2 P_H, \quad P_{0K}=\int_{\Sigma^{'}}d\sigma^2 P_K, \quad P_{0m}=\int_{\Sigma^{'}}d\sigma^2 P_m.
\end{equation}
we obtain
\begin{eqnarray}
P_m = P_{0m}\sqrt{W}+ \Pi_{m}, \quad P_K=P_{0K}\sqrt{W}+ \Pi_{K}, \quad P_H=P_{0H}\sqrt{W}+ \Pi_{H}. \label{zm}
\end{eqnarray}
Thus, it is possible to write the massive supermembrane Hamiltonian as 
 \begin{eqnarray}
    H &=& 2\pi m^2 T \alpha T_M+ 2\pi \frac{\alpha T}{T_M}\left[P_{0m}^2+P_{0K}^2+P_{0H}^2\right]+2\pi \alpha m^2 T_M \left[A^K(Z_2)-A^{K}(Z_1)\right] \nonumber \\
    & + & \frac{1}{T_M} \lim_{\epsilon\rightarrow 0} \int_{\Sigma'}dK\wedge dH \frac{1}{2}\bigg[\left(\Pi_m\right)^2+\left(\Pi_K\right)^2+\left(\Pi_H\right)^2  \nonumber \\&+&  \frac{T^2_M}{2}\{X^m,X^n\}^2 + T^2_M \ (\partial_H X^m +\{A^K,X^m\})^2+ T^2_M(m^2\partial_KX^m-\{A^H,X^m\})^2 \nonumber \\ &+&  T^2_M \ (\partial_HA^H+\{A^K,A^H\})^2+T^2_M(m\partial_KA^K+\{A^K,A^H\})^2 \nonumber \\
    & + & m^2T^2_M(\partial_KA^H)^2 + T^2_M \ T(\partial_HA^K)^2-T^2_M\{A^K,A^H\}^2  \nonumber \\ & + &  2 T^2_M\bar{\Psi}\Gamma^-\Gamma_m \{X^m,\Psi\}+2T^2_M \bar{\Psi}\Gamma^-\Gamma_K \{A^K,\Psi\}+2T^2_M\bar{\Psi}\Gamma^-\Gamma_H \{A^H,\Psi\} \nonumber \\ &+& 2 T^2_M \ T \bar{\Psi}\Gamma^-\Gamma_K\partial_H\Psi - 2T^2_M m\bar{\Psi}\Gamma^-\Gamma_H\partial_K\Psi \bigg]\label{MM},
    \end{eqnarray}
where the temporal dependence of the momenta zero modes and $A^K$ have been omitted for simplicity. This Hamiltonian possesses several distinctive features: 
First, it contains a mass term associated with the non-trivial topology of the target $LCD$ given by Eq. (\ref{fc}). This term represents the uplift of the central charge condition into this target space. It can be interpreted as a cosmological constant term associated with the presence of a non-trivial $(1,1)- knot$ in the target space $M_9\times LCD$.

Secondly, it also presents a new characteristic  term associated with the value of the single-valued tenth dynamical field, $A^K$, evaluated on the punctures.

Thirdly, it possesses non-vanishing mass terms for each of the dynamical fields, as already discussed. This last effect is responsible for  not having, at a classical level, string-like spikes with zero energy. The spectral analysis of the theory requires a proper $SU(N)$ regularization. In our analysis, we assumed that the fields are functions of $G$ and $H$ (any function of $G$ can be expressed as a function of $K$). On the punctures $G\rightarrow \pm \infty \Rightarrow K \rightarrow \pm 1$  and the dynamic fields are well defined and take a constant value. The theory is invariant under area preserving diffeomorphims of the surface which preserve the punctures. Under these conditions, we can always express the fields in terms of a basis of functions,  vanishing at the punctures, on a compact surface which is the product of the closed interval $-1 \leq K \leq 1$ and a compact space at each $K$. There always exists a measurable set of functions, the Lagrangian's eigenfunctions, which is a basis of the $L^2$ space on that compact surface. We can then implement a regularization procedure by expressing the fields in terms of the basis with coefficients depending on $t$. We may then integrate the space-like dependence and end up with a regularized version of our model.   In fact, since the quadratic terms gives mass to every field $X^m$,$A^K$ and $A^H$, and cubic contributions are bounded above by the quartic ones the spectrum associated with the bosonic Hamiltonian is expected to be purely discrete. Moreover the fermionic potential would be dominated by the bosonic potential due to the non-vanishing quadratic contribution. The Hamiltonian should then satisfy the sufficient condition found in \cite{mpgm12} and the spectrum of the supersymmetric Hamiltonian should be discrete.  We expect to provide the explicit construction of this argument elsewhere.

\subsection{Unbroken supersymmetry}
Here we will analyze the unbroken supersymmetry of the massive M2-brane on $M_9\times LCD$. In order to preserve the non-trivial cosmological term in Eq. (\ref{fc}), we need to preserve a minimal configuration characterized by

\begin{eqnarray}
   \Psi=X^m=A^K=A^H=P_m=P_K=P_H=0, \quad X^H=H, \quad X^K=K. \label{mccc}
\end{eqnarray}
Thus, this will be a supersymmetric configuration if $\delta \Psi=\delta X^M=0$ (with $M=m,H,K$). That is 

\begin{eqnarray}
   \delta X^M &=& -\Bar{\epsilon}\Gamma^M \Psi=0, \\
   \delta \Psi &=& \frac{1}{2}\Gamma^+ ((\sqrt{W})^{-1}P_M \Gamma^M+\Gamma^-)\epsilon + \frac{1}{4}\{X^M,X^N\}\Gamma^+\Gamma_{MN}\epsilon=0,
\end{eqnarray}
where $\epsilon$ is a constant spinor (see \cite{deWit2}). Now, introducing Eq. (\ref{mccc}) into the infinitesimal transformations, it is not difficult to show that
\begin{eqnarray}
   \Gamma^+\left(\Gamma^- + \frac{1}{2}\Gamma_{KH}\right)\epsilon = 0,
\end{eqnarray}
which implies that half of the supersymmetry is broken.

\subsection{Monodromies on punctured torus bundles}
Let us analyze the M2-brane on the  $\Sigma_{1,2}$, the twice punctured torus bundle.  A knot is an embedding of the topological circle $S^1$ in a 3-dimensional manifold up to continuous deformations. We consider in particular a 3-dimensional lens space $L(p,q)$ where $p$ and $q$ are relatively prime integers. These are 3-dimensional manifolds that can be built by gluing together two solid torus by a homeomorphisms of its boundary ($\Sigma$). The attaching homeomorphism wraps a meridian $p$-times latitudinally on one boundary  and $q$-times meridionally on the other boundary. In particular, $S^3$ is a lens space $L(1,0)$.  Let $H$ denote the solid torus with boundary $\Sigma$ and consider a trivial arc $A$ in its interior with end points the two points $Z_1$ and $Z_2$ on $\Sigma$.  We consider a copy of the solid torus $H$ with the arc $A$, denoted $(H^*,A^*)$ and an orientation reversing homeomorphism $h$ on $\Sigma$ preserving $Z_1$ and $Z_2$. The gluing of the two tori by the homeomorphism $h$ defines a 3-manifold with a closed knot in it. This is a $(1,1)-knot$. The construction depends only on $\Sigma_{1,2}$ and the homeomorphism $h$. The obtained closed, orientable 3-dimensional manifold is denoted $H\cup_hH^*$.

Now, we denote by $U\equiv (Z_1,Z_2)$ the set of punctures, $MCG(\Sigma_{1,2})$ the mapping class group of isotopic homeomorphisms $Y:\Sigma_{1,2} \rightarrow \Sigma_{1,2}$ with $Y(U)=U$, and by $PMCG(\Sigma_{1,2})$ the mapping class group of isotopic homeomorphisms which preserve each puncture. There is a natural epimorphism $\Omega$ from the $PMCG(\Sigma_{1,2})$ or $MCG(\Sigma_{1,2})$  to the mapping class group of $\Sigma$ which is isomorphic to $SL(2,\mathbb{Z})$

\begin{equation}
    \Omega: PMCG(\Sigma_{1,2}) \rightarrow MCG(\Sigma) \cong SL(2,\mathbb{Z})
\end{equation}
 
\begin{figure}
    \centering
    \includegraphics[scale=0.5]{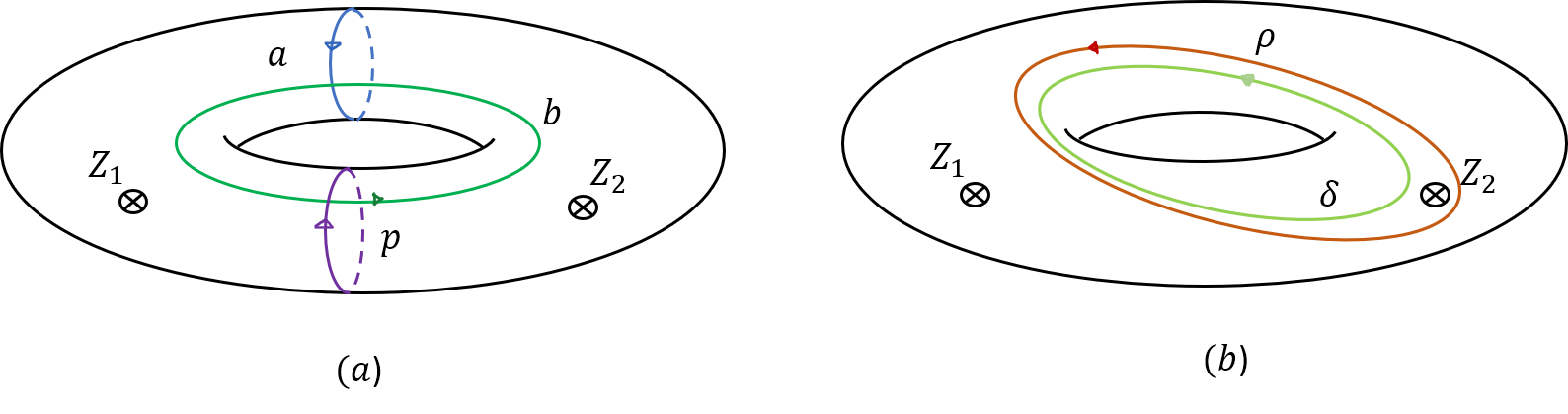}
    \caption{(a) The curves defining the Dehn twists of the generator of $PMCG(\Sigma_{1,2})$. (b) The curves associated with the generators of the ker$ \ \Omega$. }
    \label{fig:Mapping}
\end{figure}

 The Dehn twists $h_{a}$ and $h_{b}$ around the curves $a$ and $b$ associated with the basis of homology of $\Sigma$, together with the Dehn twist $h_p$ around the curve $p$ associated with the punctures are the generators of $PMCG(\Sigma_{1,2})$.            The image of $h_{a}$ and $h_{b}$ under $\Omega$ are the generators of $MCG(\Sigma)$ (see figure \ref{fig:Mapping} (a)) while $h_p$ has the same image as $h_a$ since the homeomorphisms they generate on $\Sigma$  are on the same isotopy class.  Every $(1,1)-knot$ (non-trivial knots) in a lens space $L(p,q)$ can be represented by the composition of an element in the kernel of $\Omega$ and an element depending only on the lens space (see \cite{Cattabriga}).  The Dehn twists $h_\delta$ and $h_\rho$ along the curves $\delta$ and $\rho$ in $\Sigma$, respectively (see figure \ref{fig:Mapping} (b)) can be used to construct the generators of the ker$ \ \Omega$. Specifically, the generator of ker$ \ \Omega$ are $h_m=h_ah_p^{-1}$ and $h_l=h_\rho h_{\delta}^{-1}$ where $h_{\rho}=h_m^{-1}h_{\delta}h_m$. In fact, since  $\Omega(h_a) = \Omega(h_p)$ then $\Omega(h_m)$ is the identity and so is $\Omega(h_l)$. The subgroup generated by  $h_{a}$ and $h_{b}$ generates trivial knots. There is then a surjective map between $PMCG(\Sigma_{1,2})$ and the $(1,1)-knots$.

On the torus, each isotopy class of homeomorphisms is in one to one correspondence with the isotopy classes of diffeomorphisms. Furthermore, for any dimension the full group of diffeomorphisms of a smooth manifold is homotopy equivalent to the group of volume preserving diffeomorphisms. But on a closed surface, two homotopic diffeomorphisms are also isotopic. Hence on the torus, the isotopy classes of homeomorphisms are in one to one correspondence  with the isotopy classes of area preserving diffeomorphisms, the local symmetry of the supermembrane.

The supermembrane with a punctured torus on the target space (LCD), we are proposing, can be formulated on a symplectic torus bundle with monodromy, as it was stated for the supermembrane without punctures in \cite{mpgm3}. We are interested in a torus bundle arising from a parabolic monodromy. This is essential to be able to prove Hull's conjecture concerning the origin in supermembrane theory of Roman's supergravity. Hence, if we started with a M2- brane with a parabolic monodromy, the $(p,q)$ KK charges belong to a parabolic coinvariant class. Any element of the class defines an equivalent torus bundle and the Mass operator is invariant on the class \cite{mpgm9}. The $(p,q)$ pair defines an element of the homology of the torus on the target space. Given $(p,q)$ we consider the associated lens space $L(p,q)$ and $(1,1)-knots$ on it. A relevant aspect of our construction is the effect of the punctures on the new supermembrane model. In particular, the monodromies around the punctures. However, every $(1,1)$-knot in $L(p,q)$ admits a representation consisting in the composition of an element of the group generated by the $Ker(\Omega)$ and an element which only depends on $L(p,q)$, i.e. in the original parabolic monodromy. The $(1,1)-knot$ then characterizes the monodromy construction on the punctured M2-brane we are proposing.

\section{Discussion and Conclusions} 

We have obtained the description of a massive $D=11$ supermembrane formulated on a target space locally described by $M_9\times LCD$ with monodromies, hence the theory contains ten non-compact dimensions. Because of the non-trivial topology associated with the target space, there is an induced quantization condition, reminiscent of the central charge condition. This condition has a different geometrical interpretation from the one found in \cite{Restuccia}. It does not imply the existence of a non-trivial $U(1)$ fiber bundle associated with worldvolume fluxes. Furthermore, since this condition is a direct consequence of  the punctures, we expect it to be related to the existence of brane sources at a supergravity level, specifically to the presence of M9-branes. 

The topological term, given by  Eq. (\ref{fc}), is a direct consequence of the non-trivial target space structure, and it explicitly depends on the genus one twice punctured Riemann surface. Thus, we conjecture that the topological term in our formulation is related, in the  low energy description, with the mass term of Roman’s supergravity. This is in agreement  with \cite{Bergshoeff,Sato}  where it was established a relationship between M9-branes and the cosmological constant term of  massive type IIA supergravity. The topological term is preserved only by half of the total supersymmetry. Consequently, our proposal for a massive supermembrane is invariant under half of the supersymmetric generators. This is similar to what happens in the supermembrane with central charges, in which the irreducible condition (see  \cite{Restuccia}) breaks half of the supersymmetry. Indeed, the proof is analogous to the one presented by the authors in \cite{Restuccia6}. However, as we mentioned before, the interpretation is quite different. 

The geometric structure of our proposal induces mass terms different from zero associated with all degrees of freedom of the theory. If  a supersymmetric regularization of the Hamiltonian is provided, for example on the lines argued in this paper, then the presence of mass terms in the bosonic potential together with the structure of the supersymmetric potential should  ensure a discrete spectrum. It would represent a new sector of M-theory with well-defined quantum properties. The other known cases of the supermembrane with a purely discrete spectrum correspond to: the supermembrane with central charges, the supermembrane with $C_{\pm}$ fluxes and the supermembrane on a pp-wave. Each of these cases is characterized by different topologies in the target space.

The theory is globally and locally invariant under area preserving diffeomorphisms that fix the punctures. The image of the real part of Mandelstam’s map is the whole real line $\mathbb{R}$. The punctures are mapped to   $\pm \infty$. In addition, there is a new global independent constraint around one puncture, signalling the presence of a non-trivial  monodromy around it. If the punctured torus becomes a compact torus, through a surgery procedure, this new constraint disappears.

We started with the Hamiltonian of a 11D supermembrane obtained from the compactification on a symplectic torus bundle with $C_-$ fluxes and non-trivial monodromy. In \cite{mpgm2,mpgm7} it was shown that this theory, at low energies, is described by a type II gauged supergravity with monodromies contained in $SL(2,\mathbb{Z})$. Further evidence was supported in \cite{mpgm10} by showing that the preceding theory corresponds in fact to a twisted bundle (with fluxes) formulation of the supermembrane. The theory is also U-duality invariant, and it describes a $U(1)$ Hamiltonian. It generates a  $11D$ massive supermembrane  in ten non-compact dimensions by decompatifiying the theory on a twice punctured Riemann surface and mapping it into a Light Cone Diagram $LCD$. We obtained its Hamiltonian and discussed its properties. The theory may be formulated on a bundle whose fiber is a punctured torus and whose structure group, the area preserving diffeomorphisms as suggested in \cite{mpgm13}. By using the relationship between $PMCG (\Sigma_{1,2}) $ and the $ (1,1)-knots$ theory, we show that the monodromies associated with the twice punctured torus bundle are much richer than the ones associated with torus bundles without punctures. This is because there are new monodromies characterized by the $(1,1)-knots$ constructed with the generator of $PMCG (\Sigma_{1,2}) $ while the generator of $MCG(\Sigma)$ only leads to trivial knots. Our massive supermembrane formulation is characterized by the composition of the parabolic monodromy, required for the decompactification procedure, and the monodromy around the punctures which generate the non-trivial topology contributions to the theory. 

As a final comment we would like to remark that the M2-brane contains much more degrees of freedom than the low energy effective field limit with conformal symmetries. Moreover, it does not have conformal symmetry and it is manifestly invariant under diffeomorphims. Consequently, depending on the monodromy structure of the M2-brane theory one may arrive to different low energy corners of the theory. It is also possible to consider a  conformal deformations of the theory described at low energy by the Super Chern Simons matter theory. A different explicit construction in terms of multiple M2- branes with central charges was presented in  \cite{mpgm17}.  Furthermore, in our construction, the relevant contributions arise from the monodromies around the punctures, where curvature is infinite, hence it does not satisfy the weak curvature hypothesis of \cite{Aharony2}.

It would be interesting to see whether there is a relationship with the analysis performed in \cite{Bergshoeff6} at the level of supergravity. In that study, Romans supergravity is uplifted to 11D, mainly through the existence of a 11D mass term which depends on a killing vector present in the 11D  formulation. We believe that the results presented in this work represent a concrete realization of Hull’s proposal in M-theory.
 

\section{Acknowledgements}

P.L. is supported by the Projects ANT1756 and ANT1956 of the Universidad de Antofagasta. P.L want to  thanks to CONICYT PFCHA/DOCTORADO BECAS CHILE/2019-21190517. The authors M.P.G.M. and P. L also thank to Semillero funding project SEM18-02 from U. Antofagasta, and to the international ICTP Network  NT08 for kind support.  


\end{document}